\documentclass[12pt]{article}

\usepackage{amsmath}
\usepackage{graphicx}
\usepackage{enumerate}
\usepackage{natbib}
\usepackage{url}

\usepackage{bm}
\bmdefine{\btheta}{\theta}

\begin{document}

\def\spacingset#1{\renewcommand{\baselinestretch}%
{#1}\small\normalsize} \spacingset{1}

\thispagestyle{plain}
\begin{center}
\Large{\textbf{Stable Survival Extrapolation via Transfer Learning}}
       
\vspace{0.4cm}
Anastasios Apsemidis$^{a*}$ and Nikolaos Demiris$^b$
\vspace{0.4cm}

\normalsize{$^{a*}$Department of Statistics, Athens University of Economics and Business, Athens, Greece, {\tt apsemidis@aueb.gr}} \\
\normalsize{$^b$Department of Statistics, Athens University of Economics and Business, Athens, Greece, {\tt nikos@aueb.gr}}
\end{center}

\bigskip
\begin{abstract}
The mean survival is the key ingredient of the decision process in several applications, notably in health economic evaluations. It is defined as the area under the complete survival curve, thus necessitating extrapolation of the observed data. This may be achieved in a more stable manner by borrowing long term evidence from registry and demographic data. Such borrowing can be seen as an implicit bias-variance trade-off in unseen data. In this article we employ a Bayesian mortality model and transfer its projections in order to construct the baseline population that acts as an anchor of the survival model. We then propose extrapolation methods based on flexible parametric polyhazard models which can naturally accommodate diverse shapes, including non-proportional hazards and crossing survival curves, while typically maintaining a natural interpretation. We estimate the mean survival and related estimands in three cases, namely breast cancer, cardiac arrhythmia and advanced melanoma. Specifically, we evaluate the survival disadvantage of triple-negative breast cancer cases, the efficacy of combining immunotherapy with mRNA cancer therapeutics for melanoma treatment and the suitability of implantable cardioverter defibrilators for cardiac arrhythmia. The latter is conducted in a competing risks context illustrating how working on the cause-specific hazard alone minimizes potential instability. The results suggest that the proposed approach offers a flexible, interpretable and robust approach when survival extrapolation is required.
\end{abstract}

\noindent%
\vfill

\newpage
\spacingset{1.9}
\section{Introduction}
\label{sec:intro}

Extrapolation beyond the observed survival curve can be dangerous and should not be made casually. However, practical applications necessitate such extrapolation, especially in health economic modeling and related applications of survival analysis. This is sometimes performed by fitting naive parametric models and simply projecting them to the future. Alternatively, solutions based upon the restricted mean survival (RMS) have been proposed, but from the health economic perspective these give an accurate answer to the wrong problem. 

In this paper we proposed an alternative approach building on flexible parametric survival models. We integrate several data sources from registries and population mortality along with clinical data in order to produce robust estimates of the required estimands. We proceed by incorporating mortality projections thus creating the timely stable anchor upon which extrapolation may be based. We then utilise flexible polyhazard parametric models which seamlessly incorporate the complex nature of time-to-event data, including cause-specific data and crossing survival curves.

Anchoring against external data has been recognised as a way to build towards robust ways of extrapolation to the future (see \citealt{rutherford2020nice}), either from a Bayesian perspective (e.g. \citealt{demiris2006bayesian}) or using a frequentist approach. This is essentially being proposed because it offers an implicit bias-variance trade-off in the unseen data since utilising long term information minimises the chance of wild extrapolation, albeit with the possibility of adding bias. Since this trade-off is concerned with future data, this represents a non-standard statistical problem. See \citet{jackson2017extrapolating}, \citet{bullement2023systematic} and \citet{latimer2022extrapolation} for recent reviews on survival extrapolation by anchoring against external data. In some recent developments, \citet{gallacher2021extrapolating} discuss problems regarding the goodness-of-fit statistics to select a model for extrapolation, while \citet{che2023blended} present a method that gradually blends the predicted survival of the disease population with the survival of the general population. \citet{guyot2017extrapolation} use splines to extrapolate survival of cancer patients, while the practicality of adopting this line of research has improved substantially by the approach of \citet{jackson2023survextrap} who provides an R package for flexible and practical extrapolation using M-splines.

One aspect of all these methods relates to the fact that the survival curve of the external population is based on past information of mortality. Thus, when creating the survival profile of a `synthetic' 60 year old woman, say, this effectively does not reflect the life expectancy of a concurrent individual. In this paper we adopt a different approach and use mortality projections for the external population, thus using an updated profile that appears more suitable for contemporary calculations. We then combine the external projections with richly parametric models of the poly-hazard type and capitalise on their statistical efficiency and interpretability.

Our methodology is motivated from and tested on three case studies regarding breast cancer, metastatic skin melanoma and cardiac arrhythmia. The breast cancer dataset contains long follow up and therefore facilitates the thorough assessment of the predictive ability of the proposed approach based on suitable information fractions. We use individual information on particular genes to classify individuals as triple negative or not and evaluate their life expectancy in a challenging scenario of crossing survival curves. The current developments regarding mRNA therapeutics is tackled in melanoma where we estimate the advantage when the mRNA vaccine is received along with pembrolizumab versus those receiving pembrolizumab immunotherapy alone. Our third use case is concerned with an arrhythmia dataset where we estimate the LYG for patients that receive an implantable cardioverter defibrilator (ICD) compared to a group receiving anti arrhythmia drugs (AAD). In doing so, we extend the poly-hazard models with the introduction of suitable change-points. The implementation of proposed methods has been carried out in the \citealt{rlang} language and the relevant \citealt{rstan} code is made freely available on Github.

The remaining of the article is organized as follows. Section \ref{sec:motiv} presents the three case-studies that motivated our work where survival extrapolation is of paramount importance. Section \ref{sec:extr_method} describes the methodology proposed for achieving this, while Section \ref{sec:res} summarises the results. Section \ref{sec:discuss} concludes with a discussion and ideas for future research.

\section{Motivating Problems}
\label{sec:motiv}

Here we describe the three medical applications that drive our interest in creating the class of extrapolation methods developed in this paper.

\subsection{Breast Cancer}

Breast cancer is the most common cancer indication in women in 157 out of 185 countries according to 2022 data by \citealp{who}. Therefore, accurate estimates of their survival profile is of great interest. This is true for the general cases but also for particular sub-types (e.g. \citealt{coles2024lancet}) that are increasingly of interest in `precision medicine'. Three types of genes are known to play a significant role in the survival of such patients, ER, PR and HER2. Patients with negative status in all three are typically termed `triple negative' and have a poor prognosis so their study as a group of special interest is of paramount importance (see \citealt{abraham2024partner} for example). We focus on the METABRIC breast cancer dataset (\citealt{curtis2012genomic}) which is available at \citealp{cbioportal} and include genomic and clinical data of patients from  Canada and the United Kingdom (UK). 

\subsection{Advanced Melanoma}

Melanoma refers to a type of skin cancer at metastatic stage. Their treatment has been revolutionalized due to immunotherapy such as pembrolizumab; see \citet{villani2022treatment} for a discussion of the currently available treatment options. Perhaps the most exciting new cancer treatment is based on mRNA technology and a number of clinical trials are currently being deployed, mostly on patients with resected melanoma at high risk of recurrence and is being used as an adjuvant therapy in combination with adjuvant pembrolizumab. \citet{khattak2023personalized} have published the first such trial where the hazard ratio for recurrence-free survival was 0.561 for patients receiving `mRNA-4157 (V940) + pembro' against pembro alone, while no estimates for overall survival were reported. Preliminary survival estimates of these groups are thus desirable. We digitize the published survival curve (see the on-line Appendix A) from \citet{hamid2019five} to obtain event times of patients receiving pembro. We then use the estimated hazard ratio as the basis for the `mRNA+pembro' curve. In the on-line Appendix F, we gather the results from a more recent clinical trial (\citealt{robert2023seven}).

\subsection{Cardiac Arrhythmia}

Cardiac arrhythmia refers to problems regarding the rate of the heartbeat. Survivors of ventricular fibrillation receive AAD treatment to control the disease, but ICD was shown to be more efficient (see \citealt{connolly2000meta}). We aim to estimate the LYG among patients on AAD or with ICD in a cause-specific hazards setting. We digitized the published Kaplan-Meier curve from Demiris and Sharples (2006) and use the meta-analysis of \citet{connolly2000meta} who report a hazard ratio $h_{ICD} / h_{AAD}=0.5$ with 95\% C.I. $(0.37,0.67)$ for arrhythmia-related deaths, see also \citet{buxton2006review}. The `death due to other causes' component is common in both groups and the external population thus minimizes the risk of unstable extrapolation.

\section{Extrapolation using mortality projections}
\label{sec:extr_method}

\subsection{Polyhazard models}

Here we define the building block of our approach, a flexible parametric survival model. Let us set the notation and denote by $h(t):=\displaystyle\lim_{h\rightarrow 0}\frac{P(t\leq T < t+h \,|\, T\geq t)}{h}$ the hazard at time t and $S(t)$ the survival function. Then, given data ($t_i, \delta_i$), $i=1,\dots,n$ with $t_i$ the observed time and $\delta_i$ the corresponding indicator of either event or censoring, the likelihood function reads $L = \prod_{i=1}^n S(t_i) h(t_i)^{\delta_i}$. The proposed models are best modeled and assessed at the hazard scale. We build on their empirical estimates and in addition to the Kaplan-Meier estimator (\citealt{kaplan1958nonparametric}) we use empirical hazard estimates, e.g. \citet{muller1994hazard} and \citet{gefeller1992nearest}. Poly-hazard models (see for instance \citealt{louzada1999polyhazard} and \citealt{demiris2015survival}) postulate that the hazard function $h(\cdot)$ can be decomposed as a sum of $M$ components: $h(t)=\sum_{m=1}^Mh_m(t)$.

We define a joint model for the external and disease population, both of which are assigned a poly-hazard process. We denote by $h_d(t)=\sum_{m=1}^Mh_d^m(t)$ the hazard of the disease group and $h_p(t)=\sum_{m=1}^Md_p^m(t)$ the hazard of the external population data and assume that some of their components are either identical or proportional to each other. A selection that worked well in our case studies uses $M=3$, $h_d^1(t)=C\cdot h_p^1(t)$ and $h_d^3(t)=h_p^3(t)$. Finally, the likelihood contributions from the two datasets are
\begin{align}
L_d &= \prod_{i=1}^{n_d}h_d(t_i)^{\delta_i}S_d(t_i) \quad \mathrm{and} \label{eq:diseaselike} \\
L_p &= \prod_{j=1}^{n_p}h_p(t_j)^{\delta_j}S_p(t_j) \label{eq:poplike}
\end{align}
for samples of size $n_d$ and $n_p$ respectively, composing the total likelihood
\begin{equation} \label{eq:totlike}
L=L_d\cdot L_p
\end{equation}

Selecting the parametric form of each hazard component represents a model determination issue. The Weibull family has been used extensively due to its combination of tractability and flexibility, including increasing, constant and decreasing hazards. Thus, the poly-Weibull model can result in the commonly observed bathtub (or U-shaped) hazard, in contrast to mixtures of Weibull densities (\citealt{glaser1980bathtub}). The Log-Normal and Log-Logistic-based hazards can be either decreasing or unimodal and thus capable of producing an upside-down U-shaped hazard. They also represent resonable options for increasing components since the maximum may be reached far beyond the observed time period. Note that there is no restriction in using single-family components and mixing is recommended if inspection of the empirical hazard suggests this.

\subsubsection{Polyhazard models with change-points}

One may add flexibility by incorporating change-points to the hazard rate, e.g. \citet{friedman1982piecewise} or \citet{casellas2007bayesian}. We extend this approach to the polyhazard case. Suppose there exist $K$ fixed change-points in time $\tau_{0:K-1}$, where the total hazard adopts a different form of a polyhazard model with parameters $\btheta_k$, i.e. $h(t; \btheta)=h(t;\btheta_k)I\big(t\in (\tau_{k-1},\tau_k)\big)$, for instance a Bi-Weibull form $h_W(t;\alpha_1,\lambda_1)+h_W(t;\alpha_2,\lambda_2)$ in the interval $(\tau_0, \tau_1-1)$ and a Tri-LogNormal form $h_{LN}(t;\mu_1,\sigma_1)+h_{LN}(t;\mu_2,\sigma_2)+h_{LN}(t;\mu_3,\sigma_3)$ in the $(\tau_1, \tau_2)$ interval, where $\tau_0=0$ and $\tau_2=max_{i}\{\mathbf{t}_{t_1,...,t_n}\}$. Then, the survival function may be written as
\begin{equation*}
S(t;\btheta)=
\left\{\begin{alignedat}{2}
&S(t;\btheta_1) &,\quad& 0 \leq t < \tau_1 \\
&S(\tau_1;\btheta_1)S(t;\btheta_2)/S(\tau_1;\btheta_2) &,\quad& \tau_1 \leq t < \tau_2 \\
&... \\
&S(\tau_1;\btheta_1)\displaystyle\frac{S(\tau_2;\btheta_2)}{S(\tau_1;\btheta_2)}\cdot...\cdot\displaystyle\frac{S(t;\btheta_K)}{S(\tau_{K-1};\btheta_K)} &,\quad& \tau_{K-1} \leq t \leq \tau_K \\
\end{alignedat}\right.
\end{equation*}
or compactly as $S(t;\btheta)=S(t;\btheta_J)\displaystyle\frac{\prod_{j=1}^{J-1}S(\tau_j;\btheta_j)}{\prod_{j=1}^{J-1}S(\tau_j; \btheta_{j+1})} I\big(t\in (\tau_{J-1}, \tau_J)\big)$, i.e. appropriately adjusting the survival function at each change-point. The likelihood still reads $L=\prod_{i=1}^nh(t_i;\btheta)^{\delta_i}S(t_i;\btheta)$ using the amended hazard and survival functions. An alternative way to visualize this formula is by denoting with $s\big((a,b)\big)$ the survival of a patient inside the interval $(a,b)$. Then the survival function reads $S(t)=s\big((0,\tau_1)\big)s\big((\tau_1,\tau_2)\big)...s\big((t,\tau_K)\big)$ where each patient may reach time $t$ after surviving each interval of length $\tau_1$, $\tau_2-\tau_1$, $...$, $\tau_K-t$ with parameters $\btheta_1$, $\btheta_2$, $...$, $\btheta_K$ respectively.

\subsection{Construction of the external population via mortality projections}

Several authors argued that projecting survival beyond the last observed time $t^*$ should anchor against external long-term evidence. This makes intuitive sense since the long-term data facilitate adding stability and may prevent wild extrapolations. Such external survival data have typically been based on past survival information as opposed to current estimates of life expectancy. We amend this issue by adopting projections of the mortality rate, thus creating estimates of the current life expectancy for the external data. In this paper these projections are based on the Lee-Carter model (\citealt{lee1992modeling}) but alternative models may be used. Denote by $y_{x,t}$ the log-mortality rates for age $x$ and time $t$ and write
\begin{align*}
y_{x,t}&=\alpha_x+\beta_x\kappa_t+\epsilon_{x,t} \\
\kappa_t&=u+\kappa_{t-1}+v_t
\end{align*}
where $\epsilon_{x,t}\sim N(0,\sigma_\epsilon^2)$ and $v_t\sim N(0,\sigma_v^2)$. We follow the Bayesian approach of \citet{pedroza2006bayesian} and project the mortality ${}_{t'}m_{x}$ for age $x$ at a future time $t'$. The probability of reaching age $x+j$ from age $x$ is estimated as $${}_{t'}\pi_x^{x+j}=\exp[-\sum_{i=0}^{j-1} {}_{t'}m_{x+i}]$$ thus providing an estimate of the survival probability for the external population at the desired time. For instance, the probability a 20-year-old survives until the age of 22 in 2024 is estimated as ${}_{2024}\pi_{20}^{22}=\exp[-{}_{2024}m_{20}-{}_{2024}m_{21}]$. Using this approach one may create synthetic times until death from the external population provided it has been age-sex matched with the demographic characteristics of the clinical data. The mortality data can be obtained from the \citealp{hmd}. For example, we downloaded UK mortality rates for ages 0-110 of years 2000-2020, we projected until 2023 and we utilized the published proportions of arrhythmia-related deaths at each age in order to generate the synthesized data as in \citet{benaglia2015survival}.

\subsection{Extrapolation methods}

The building block for moving beyond the observed time horizon is the joint poly-hazard model and the corresponding estimates of the two total hazards, for the disease group, $h_d(t;\btheta_d)$, and for the general population $h_p(t;\btheta_p)$. We proceed by proposing two methods for obtaining the required extrapolated curves. The basic method projects $h_d(t;\btheta_d)$ using the estimated parameters $\hat{\btheta}_d$. Note this approach implicitly uses the external information via its dependence on the components of $h_p$ retaining the required robustness. An alternative approach for extrapolating $h_d$ is postulating that at the end of the follow-up period the relationship between $h_d$ and $h_p$ remains unchanged in some form. Specifically, one may assume an additive relation and calculate the last $k$ differences between $h_d$ and $h_p$, use the average, say $D=\frac{1}{k}\sum_{i=1}^kh_d(t_{n_d-i+1})-h_p(t_{n_d-i+1})$ and utilise $h_d(\mathbf{t}')=D+h_p(\mathbf{t}')$. Alternatively, a multiplicative relation may be employed through a (constant) hazard ratio $R=\frac{1}{k}\sum_{i=1}^kh_p(t_{n_d-i+1})/h_p(t_{n_d-i+1})$, so that $h_d(\mathbf{t}')=R\cdot h_p(\mathbf{t}')$.

The individual components of the poly-hazard models may or may not have a natural interpretation. However, one can naturally think of the component that induces high hazard at the start of follow-up as associated with the disease of interest and the component that increases as the one capturing the aging/long term effects. In that case one may apply the constant difference/ratio method for the component of interest alone and then add the extrapolated components to obtain the total hazard in this pseudo-cause-specific hazard scenario. In the Appendix we give examples illustrating each proposed method termed `baseline', `constant difference', `constant ratio', `pseudo cause specific constant difference' and `pseudo cause specific constant ratio'.

\subsection{Cause-specific hazards}

In the presence of cause-specific evidence one may resort to cause-specific hazards as follows, see also \citealt{benaglia2015survival}. Assume that the external population hazard consists of two components from the same family of distributions, one of which relates to the cause of interest and the other refers to all other causes. For component $k=1,2$ we have failure times $t_i^k$ and their corresponding censoring indicator $\delta_i^k$, $i=1,...,n_k$ and the likelihood contribution of the `cause of interest' and `other causes' is $L_p^k=\prod_{i=1}^{n_k}h_p^k(t_i^k)^{\delta_i^k}S_p^k(t_i^k)$ whence the total hazard reads $ h_p=h_p^1+h_p^2$. This decomposition means that the hazard of death from `other causes' may reasonably be assumed to be common. This facilitates for stable extrapolation since one then needs only to model the cause-specific hazard. Assuming it is proportional to the hazard of the external population we get
\begin{equation*}
h_d(t)=C\cdot h_p^1(t)+h_p^2(t)
\end{equation*}
and the likelihood reads $L_s=\prod_{i=1}^n h_d(t_i)^{\delta_i}S_d(t_i)$. The joint likelihood of the external population and the disease group is then $L=L_s\cdot L_p^1 \cdot L_p^2$.

\subsection{Mean survival and life years gained}

The mean survival is defined as the area under the complete survival curve. The RMS is given by the area under the survival curve from time $t=0$ until the end of follow-up period and this can be very far from the true mean survival in specific applications such trials without extensive follow-up or diseases with low mortality. Once the complete survival curve has been estimated one may derive the mean survival using, say, the trapezoidal rule
\begin{equation} \label{eq:traprule}
\int_0^{t_{max}}S(t)dt \approx \sum_{z=1}^N \frac{S(t_{z-1})+S(t_z)}{2}(t_z-t_{z-1})
\end{equation}
where $t_{max}$ is the maximum time point when survival probability is effectively zero and $\{t_z \,|\, z=0,...,N\}$ is a partition of the interval $(0,t_{max})$ with $t_0=0$ and $t_{max}=t_N$. The LYG between the survival functions $S_1(\cdot)$ and $S_2(\cdot)$ can then be estimated as
\begin{equation} \label{eq:lyg}
LYG_{S_1,S_2}=\int_0^{t_{max}}S_1(t)dt-\int_0^{t_{max}}S_2(t)dt
\end{equation} 
where $t_{max}=max\{t_{max}^1,t_{max}^2\}$.

\subsection{Priors and Computation}

We use the NUTS (\citealt{hoffman2014no}) version of Hamiltonian Monte Carlo as implemented in the \citealt{rstan} probabilistic programming language. We found it has been sufficiently fast and therefore resorting to other approximations such as variational inference appeared unnecessary. We list here the prior densities assigned to the main parameters. For the shape parameters of the poly-hazard models we use Exponential priors with mean 10 and for the rate parameters we use Gamma(2,0.5) priors. The proportionality constant between components of the external population and the disease group was also assigned an Exponential prior density. For the poly-LogNormal models we use Gaussian $N(0,5^2)$ priors for the location parameters.

\section{Experiments}
\label{sec:res}

The poly-Weibull model has been our baseline approach due to its flexibility, interpretability and fidelity to the data. Poly-LogNormal and poly-LogLogistic models have also been used achieving generally worse or comparable fit. Two or three hazard components have been proven sufficient in all our applications, irrespective of the selected hazard form. Here we outline the main results while additional findings are in Appendix C for the external population mortality projections, in Appendix D for the triple-negative and non-triple-negative groups of breast cancer and in Appendices E, F and G for the breast cancer, melanoma and arrhythmia data respectively.

\subsection{Breast Cancer}

\begin{figure}[h!]
\begin{center}
\includegraphics[width=0.495\textwidth]{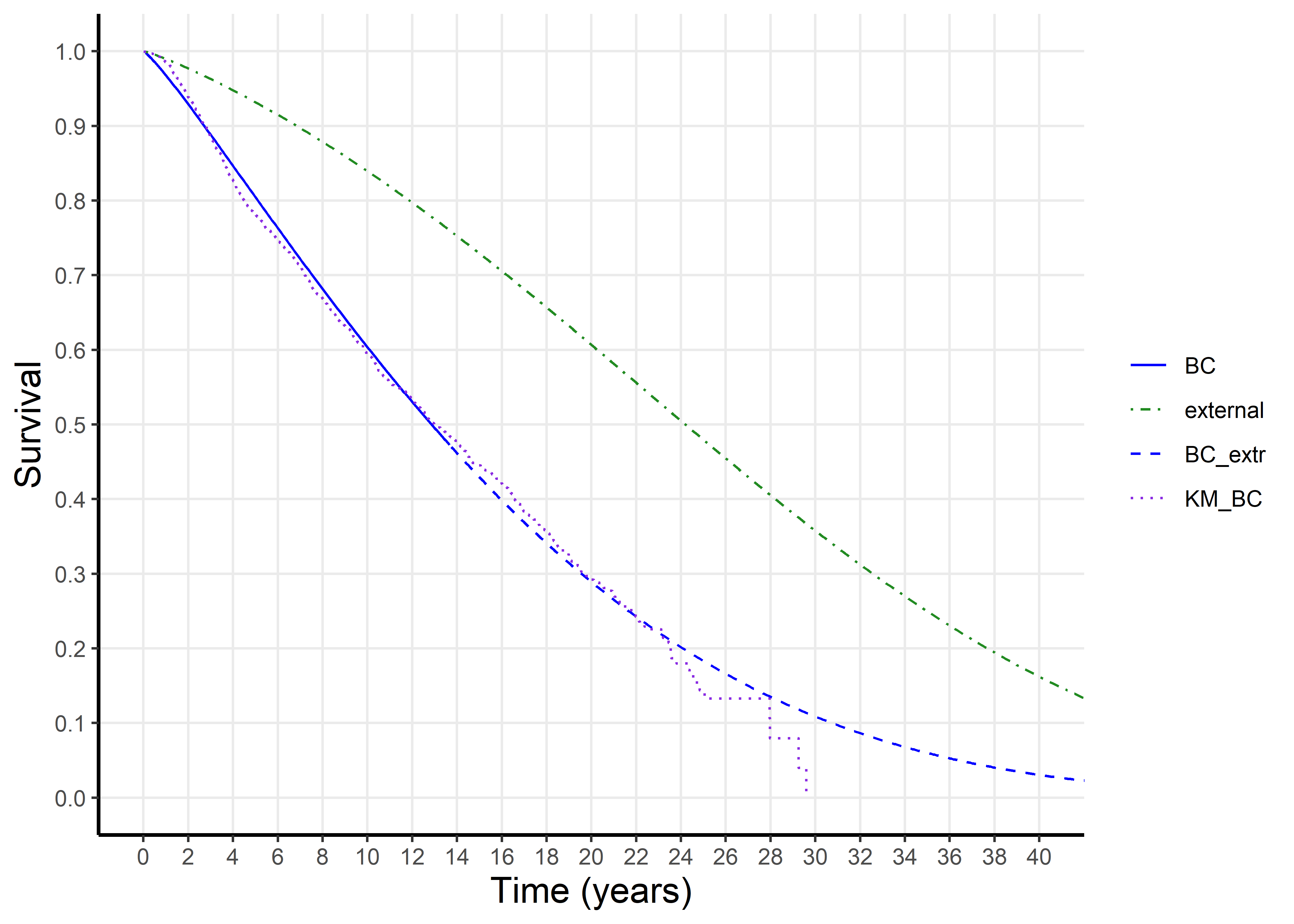}
\includegraphics[width=0.495\textwidth]{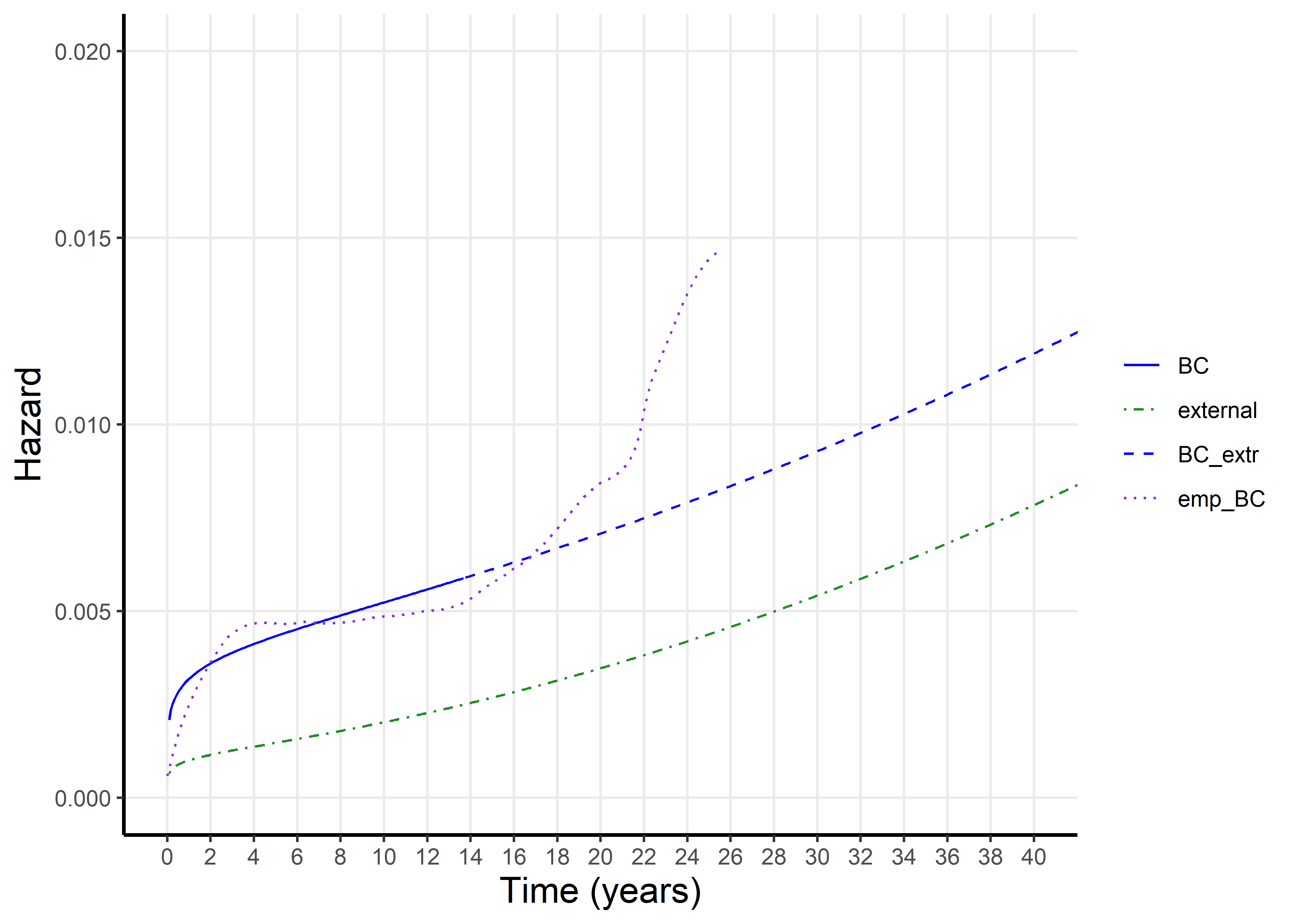}
\end{center}
\caption{Extrapolated survival function (left) and hazard function (right) when keeping only 80\% of fractional information for the breast cancer data. The breast cancer follow-up is indicated as `BC' in solid line and its extrapolation (`$\mathrm{BC_{extr}}$') in dashed line. The external population is indicated with a dot-dashed line and the empirical estimation (the KM in the survival case) as a dotted line.}
\label{fig:bc_haz_surv}
\end{figure}

The breast cancer dataset has long follow up and hence a nearly complete survival curve, thus facilitating repeated testing of the trained models against the `truth' in the form of the observed data. We use various levels of the information fraction (\citealt{demets1994interim}) as a metric. For example, 80\% of information corresponds to 49\% survival probability. The best model for this dataset was a joint Bi-Weibull for the disease group and the external population. We impose the constraint that the first components of each Bi-Weibull model will be proportional to each other and the second components will be equal. Thus, we generate age-sex matched data from the external population $y_{1:n_p}$ and we assume each sample $i=1,...,n_p$ has hazard $h_p(y_i; \alpha_1, \alpha_2, \lambda_1, \lambda_2)=h_p^1(y_i; \alpha_1,\lambda_1)+h_p^2(y_i; \alpha_2,\lambda_2)$. The hazard form of each component $m$ is Weibull with shape and rate parameters $\alpha_m$ and $\lambda_m$ respectively, i.e. $h_p^m(y_i; \alpha_m, \lambda_m)=\lambda_m \cdot \alpha_m \cdot y_i^{\alpha_m-1}$. The survival function is $S(y_i; \alpha_1, \alpha_2, \lambda_1, \lambda_2)=S_p^1(y_i; \alpha_1,\lambda_1) \cdot S_p^2(y_i; \alpha_1,\lambda_1)$, where each component takes the form $S_p^m(y_i; \alpha_m, \lambda_m)= \exp(- \lambda_m \cdot y_i^{\alpha_m})$. The likelihood is given by equation (\ref{eq:poplike}). For the disease data $t_{1:n_d}$, we also assume a Bi-Weibull model, but every sample $i=1,...,n_d$ has hazard $h_d(t_i; \alpha_1, \alpha_2, \lambda_1, \lambda_2)=C \cdot h_p^1(t_i; \alpha_1,\lambda_1)+h_p^2(t_i; \alpha_2,\lambda_2)$ and survival $S(t_i; \alpha_1, \alpha_2, \lambda_1, \lambda_2)=S_p^1(t_i; \alpha_1,\lambda_1)^C \cdot S_p^2(t_i; \alpha_1,\lambda_1)$, where $C>0$. The likelihood is given by (\ref{eq:diseaselike}) and the complete likelihood by (\ref{eq:totlike}).

The extrapolated survival and hazard functions are depicted in Figure \ref{fig:bc_haz_surv}. In this example the five extrapolation methods give similar results and we present the baseline method which in this case coincides with the pseudo cause-specific constant ratio method (by construction, see Appendix B). Thus, we use $h_d(t_i';\hat{\alpha_1}, \hat{\lambda_1}, \hat{\alpha_2}, \hat{\lambda_2})$ for each new point $t_i'$. Using equations (\ref{eq:traprule}) and (\ref{eq:lyg}), we estimate a mean survival of approximately 180 months in comparison to the external mean survival of approximately 302 months, i.e. 122 life months lost on average due to cancer. The other extrapolation methods give similar results, except of the constant ratio method estimating approximately 173 months of mean survival, slightly closer to 168, the Kaplan-Meier-based estimate. However, due to a small sample at the the of the follow-up we tend to believe the aforementioned model more than the non-parametric estimate of the KM.

\subsubsection{Subgroup Analysis for precision medicine}

\begin{figure}[h!]
\begin{center}
\includegraphics[width=0.85\textwidth]{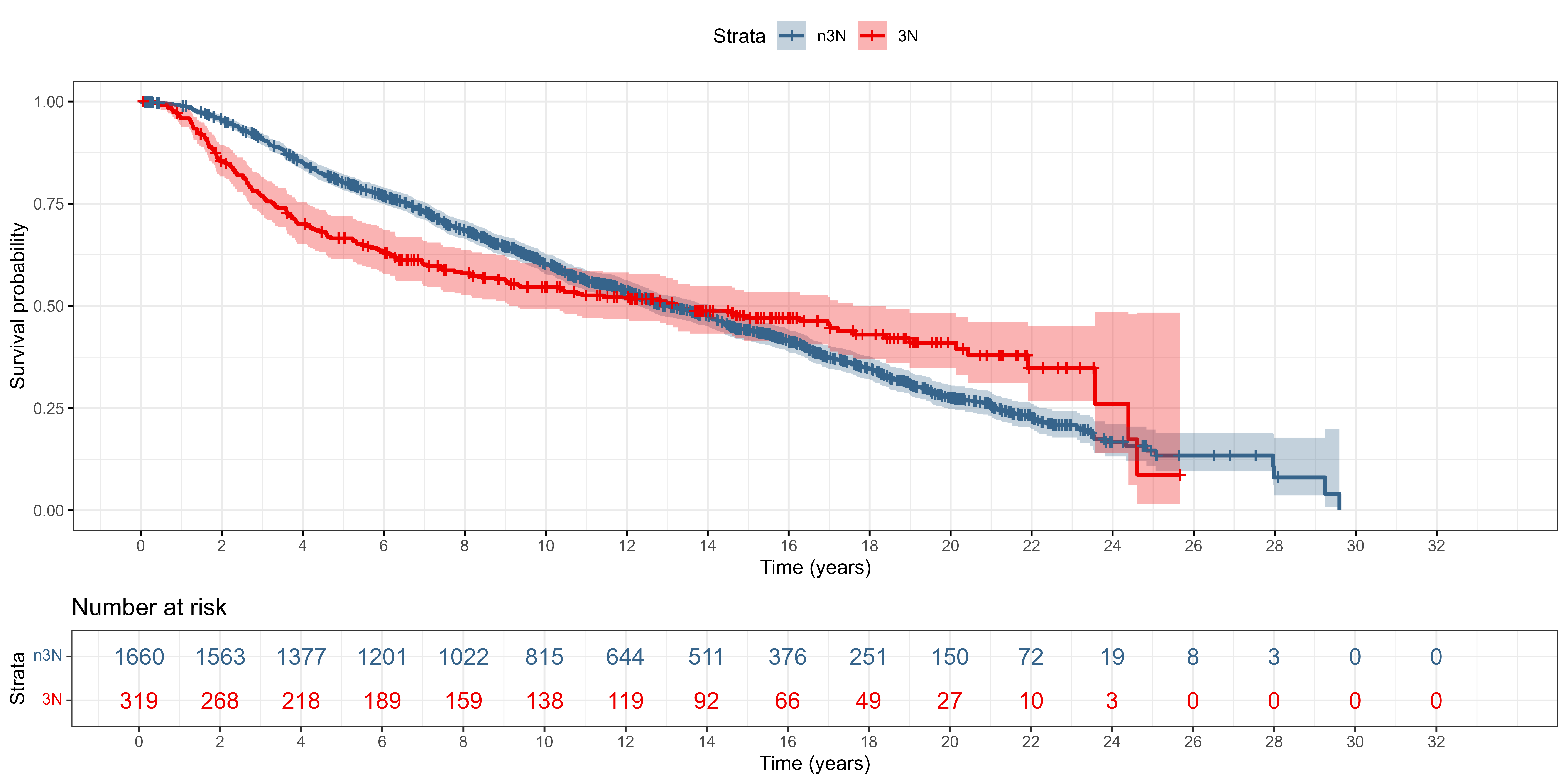}
\end{center}
\caption{The crossing survival curves of the 3N versus n3N groups.}
\label{fig:bc_kmgroups_pub}
\end{figure}

In the same dataset the status of the three genes ER, PR and HER2 are available, so we construct the groups 3N (negative to all three) and n3N (at least one positive) and try to estimate the mean survival time of each group. The 3N group is known to have a worse prognosis and such sub-grouping is becoming increasingly important in precision medicine. These two groups for the present dataset result in crossing survival curves (Figure \ref{fig:bc_kmgroups_pub}) and therefore non-proportional hazards, a challenging modelling task. However, we apply the proposed framework and naturally capture the desired data characteristics. We fit two, three, or four Weibull, LogLogistic or LogNormal components and also combinations of Weibull and LogNormal components and conclude on a joint Tri-LogLogistic model for the 3N, n3N groups and the external population with common 3rd component. This model estimates the mean survival time of 3N as 12.6 years (RMS was 13.5 years) and that of n3N as 14 years (RMS was 14.1) and thus the 3N group on average has approximately 17 months of life less compared to n3N. 

We also fitted models with data-informed change-points at 3 and 16 years, namely a Bi-LogLogistic model on the 3N group and a joint Tri-LogLogistic for the external population and the n3N group with common 3rd component but those models did not capture the data features as well, particularly the sigmoid-type shape. The different models are compared in Table \ref{t:infcrit} while the relevant figures can be found in the Appendix.

\begin{table}
\caption{Information criteria for the 3N group. Model 1 is an Exponential-Weibull-Weibull for the three parts, while Model 2 is a Bi-Weibull in each part. The results are rounded to 2 decimal digits. DIC\textsubscript{2} is the DIC version that uses the variance of the log-likelihood.}
\label{t:infcrit}
\begin{center}
\begin{tabular}{ c c c c c c c c c c c} 
\hline
& AIC & BIC & DIC & DIC\textsubscript{2} & WAIC & Time $(s)$ & $p_d$ & $p_v$ & $p_w$ & $p$ \\ \hline
Model 1 & 2122.25 & 2141.08 & 2117.98 & 2118.56 & 2117.95 & 25 & 2.86 & 3.14 & 2.8 & 5 \\
Model 2 & 2110.36 & 2155.54 & 2095.22 & 2095.93 & 2094.37 & 355 & 4.43 & 4.78 & 3.54 & 12 \\
\hline
\end{tabular}
\end{center}
\end{table}

\subsection{The effect of mRNA therapeutic on Melanoma}

\begin{figure}[h!]
\begin{center}
\includegraphics[width=0.495\textwidth]{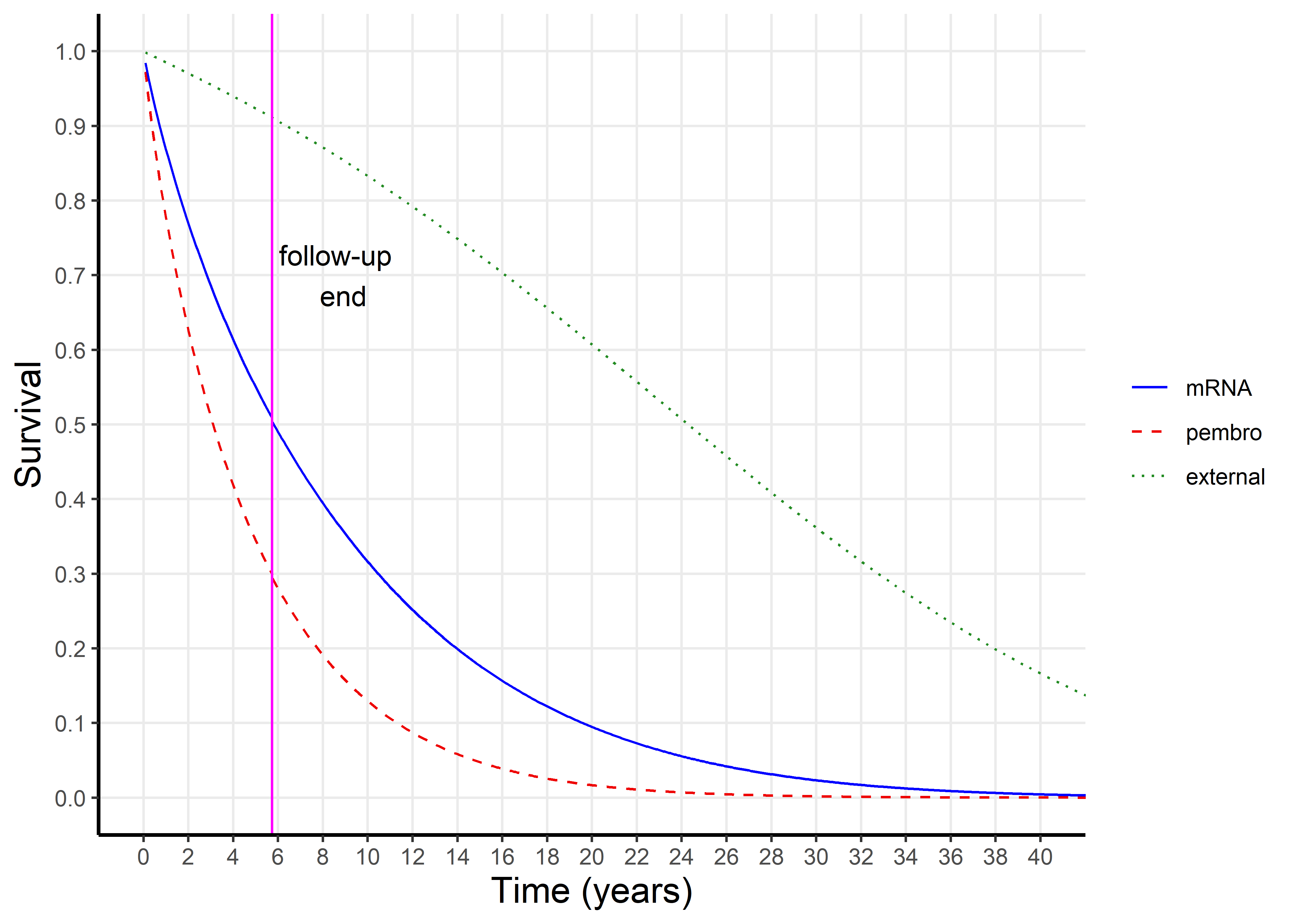}
\includegraphics[width=0.495\textwidth]{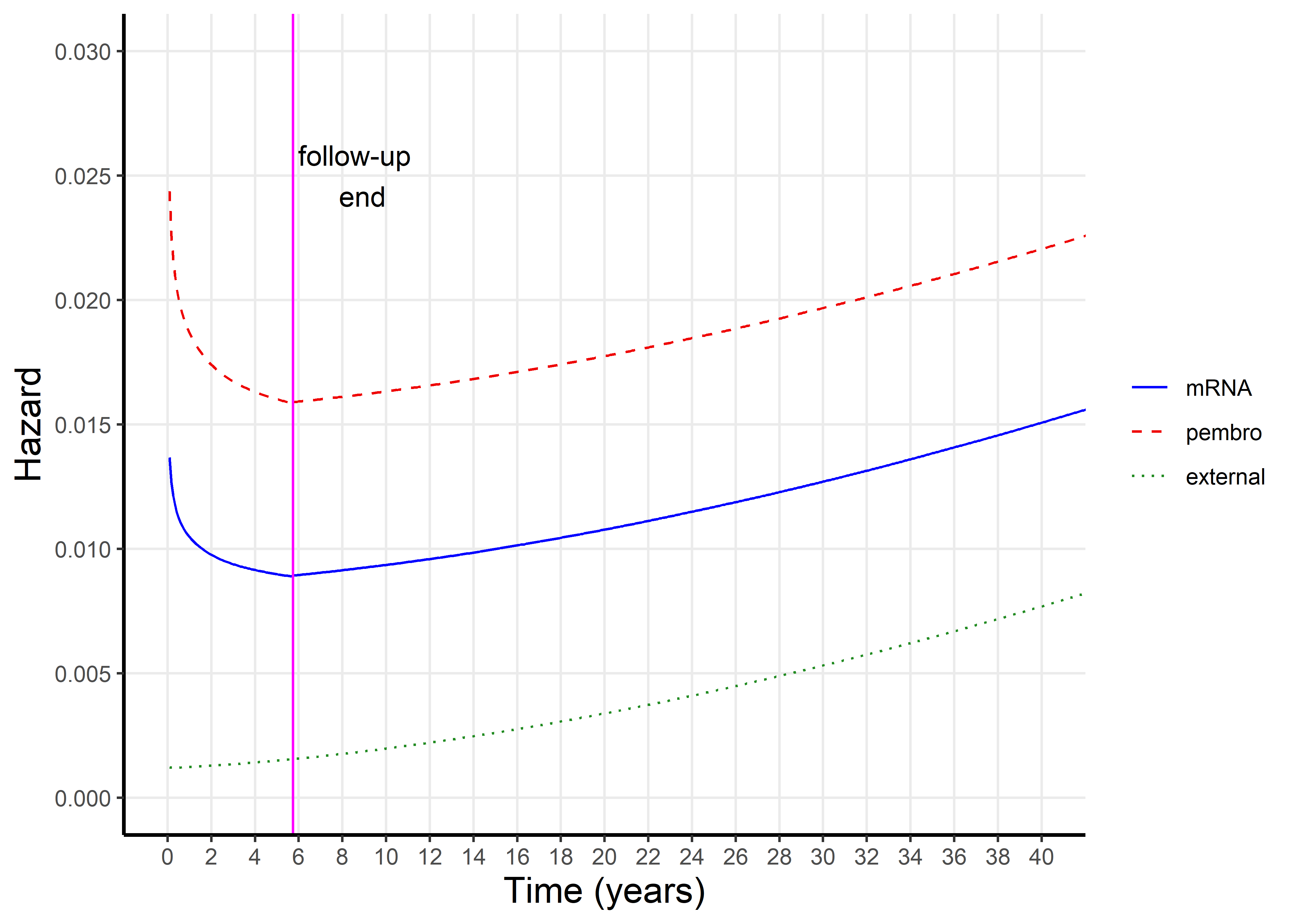}
\end{center}
\caption{Extrapolated survival function (left) and hazard function (right) for the melanoma data. We indicate the only-pembrolizumab treated group as ``pembro'' (dashed line) and the pembrolizumab plus mRNA vaccine as ``mRNA'' (solid line). The external population is indicated as dotted line and the end time of follow-up is indicated by a vertical solid line.}
\label{fig:mel_haz_surv}
\end{figure}

This treatment is used as an adjuvant therapy in combination with adjuvant pembrolizumab in patients with resected melanoma at high risk of recurrence. The best fitted model for the overall survival in the advanced melanoma dataset was a joint Tri-Weibull where the first components of the disease and external groups are proportional to each other while the third components are equal. Thus, we use $m=3$ components: $h_p(y_i; \alpha_1, \alpha_2, \alpha_3, \lambda_1, \lambda_2, \lambda_3)=h_p^1(y_i; \alpha_1,\lambda_1)+h_p^2(y_i; \alpha_2,\lambda_2) + h_p^3(y_i; \alpha_3,\lambda_3)$ and the hazard of the pembrolizumab group is given by $h_{pembro}(t_i; \alpha_1, \alpha_d, \alpha_3, \lambda_1, \lambda_d, \lambda_3)=C \cdot h_p^1(t_i; \alpha_1,\lambda_1) + h_d^2(t_i;\alpha_d,\lambda_d) + h_p(t_i;\alpha_3,\lambda_3)$. In this example we select the `constant difference' method and use the last 5 points (months) to write $h_{pembro}(t')=D_{pembro}+h_p(t')$ for a future time $t'$. The mRNA group of patients is derived via the proportional hazards assumption $h_{mRNA}(t)=0.561\cdot h_{pembro}(t)$ (see \citealt{khattak2023personalized}) where we assume that the reported hazard ratio (HR) may reasonably approximate the HR for overall survival. Extrapolation follows using $h_{mRNA}(t')=D_{mRNA}+h_p(t')$ for future times $t'$.

\begin{figure}[h!]
\begin{center}
\includegraphics[width=\textwidth]{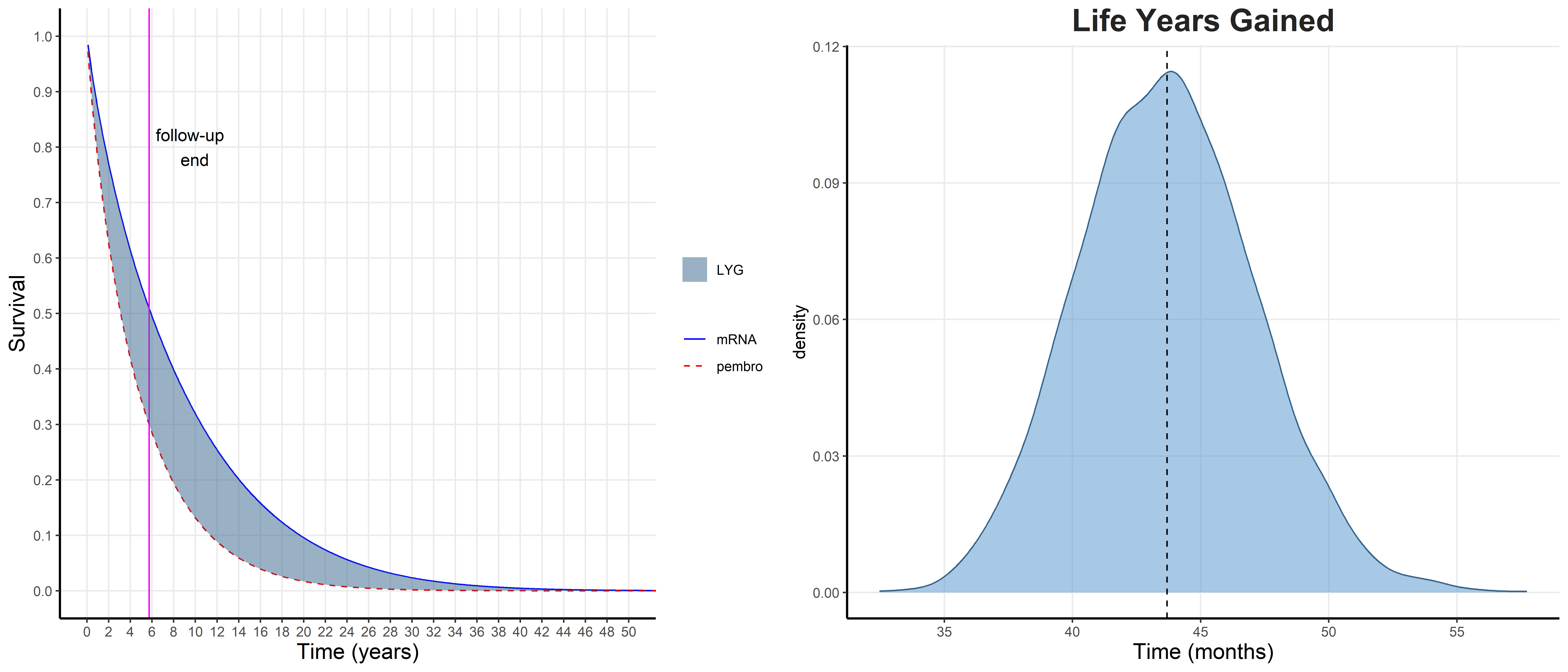}
\end{center}
\caption{LYG for the melanoma data. Left: LYG illustrated as the area between the two survival curves with the end time of follow-up noted by a vertical line. Right: Density of the LYG with its mean indicated by a vertical line.}
\label{fig:mel_lyg}
\end{figure}

Figure \ref{fig:mel_haz_surv} depicts the extrapolated survival curves and hazard functions. Calculating the area between the mRNA and pembrolizumab curves estimates that the adjuvant mRNA therapeutic offers on average an additional 43.69 months of life (see right plot of Figure \ref{fig:mel_lyg}). One may broadly interpret this as patients with advanced melanoma may gain 3.64 years of life if the mRNA therapeutic is added to their treatment. In the absence of HR estimates for overall survival we used the corresponding HR estimate for recurrence free survival and suitable sensitivity analyses ought to be adopted if these estimates are to be used for decision support/making.

\subsection{Cardiac Arrhythmia}

\begin{figure}[h!]
\begin{center}
\includegraphics[width=0.495\textwidth]{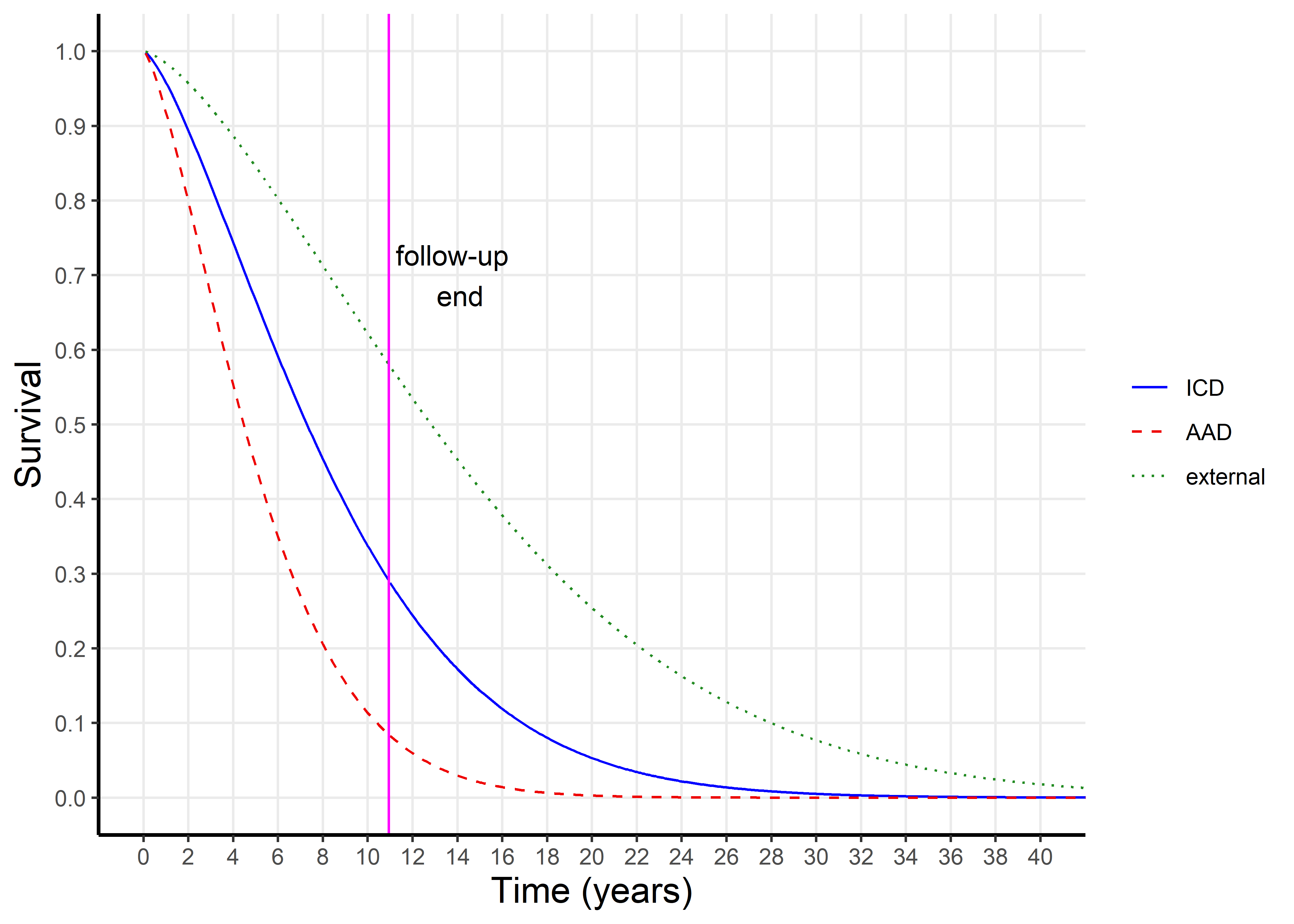}
\includegraphics[width=0.495\textwidth]{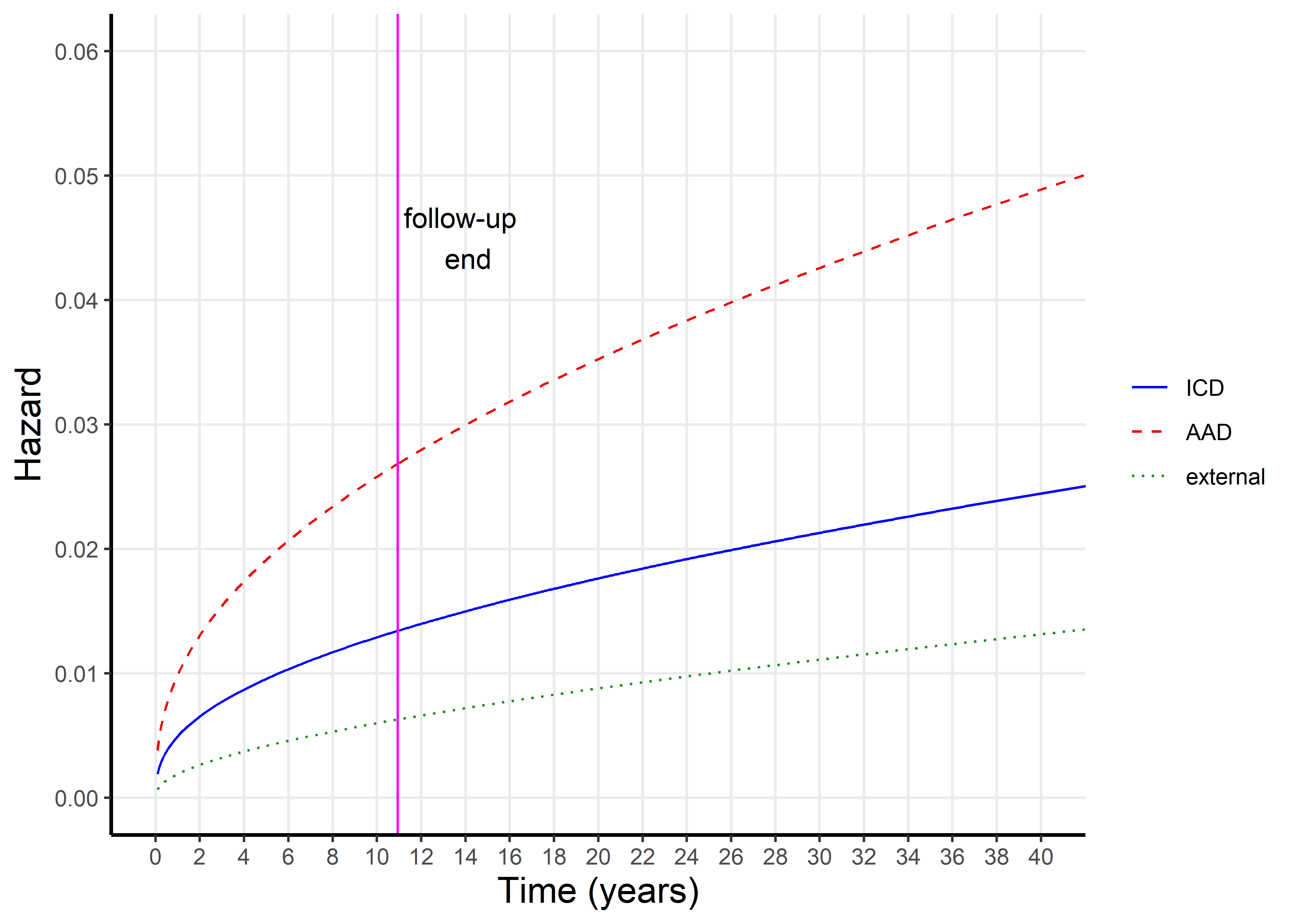}
\end{center}
\caption{Extrapolated survival function (left) and hazard function (right) for the arrhythmia data. The external population is indicated as dotted line and the end time of follow-up is indicated by a vertical solid line.}
\label{fig:ca_haz_surv}
\end{figure}

For this dataset  we focus on the Bi-Weibull model following \citet{benaglia2015survival} who found it preferable compared to a Bi-Gompertz distribution. We assume that in the external population the times until death due to arrhythmia, $y_{1:n_{p_1}}$, have a Weibull density with hazard $h_p^1(y_i; \alpha_1, \lambda_1)=\lambda_1 \cdot \alpha_1 \cdot y_i^{\alpha_1-1}$ and the times until death from other causes, $z_{1:n_{p_2}}$, a Weibull with hazard $h_p^2(y_i; \alpha_2, \lambda_2)=\lambda_2 \cdot \alpha_2 \cdot z_i^{\alpha_2-1}$. Then, the joint likelihood is given by:
\begin{equation*}
L_p(\alpha_1,\alpha_2,\lambda_1,\lambda_2)=\prod_{i=1}^{n_{p_1}} h_p^1(y_i)^{\delta_i^1} S_p^1(y_i)\prod_{j=1}^{n_{p_2}} h_p^2(z_j)^{\delta_j^2} S_p^2(z_j)
\end{equation*}
Since the cause of death is unknown in the disease group data $t_{1:n_d}$ we model them using a Bi-Weibull model with likelihood
\begin{equation*}
L_d(C,\alpha_1,\alpha_2,\lambda_1,\lambda_2)=\prod_{i=1}^{n_d} h_d(t_i)^{\delta_i} S_d(t_i)
\end{equation*}
where $h_d=C\cdot h_p^1+h_p^2$ and $S_d=(S_p^1)^C\cdot S_p^2$.

We focus upon the baseline extrapolation method which in this example coincides with the cause specific constant ratio method. The AAD group is constructed using the published hazard ratio of 0.5 (see Section \ref{sec:motiv}) for the arrhythmia-related events so that $h_{ICD}(t)=0.5\cdot h_{AAD}(t)$. Figure \ref{fig:ca_haz_surv} depicts the extrapolated survival and hazard functions. The estimated LYG are approximately 39.7 months, or 3.31 years on average (see right plot of Figure \ref{fig:mel_lyg}) since AAD and ICD patients have a mean survival of 62.27 and 101.97 months respectively.

\begin{figure}[h!]
\begin{center}
\includegraphics[width=0.495\textwidth]{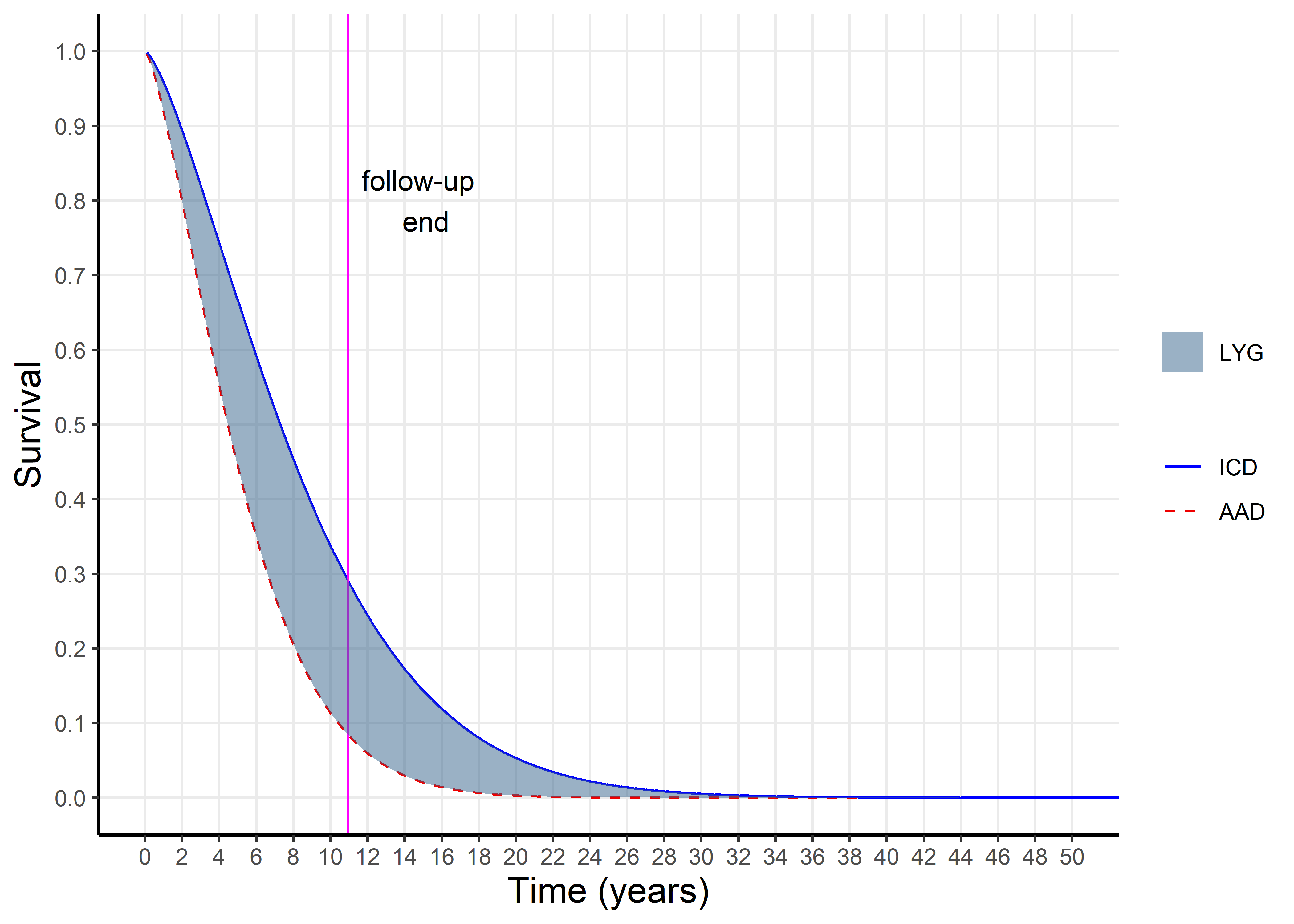}
\includegraphics[width=0.495\textwidth]{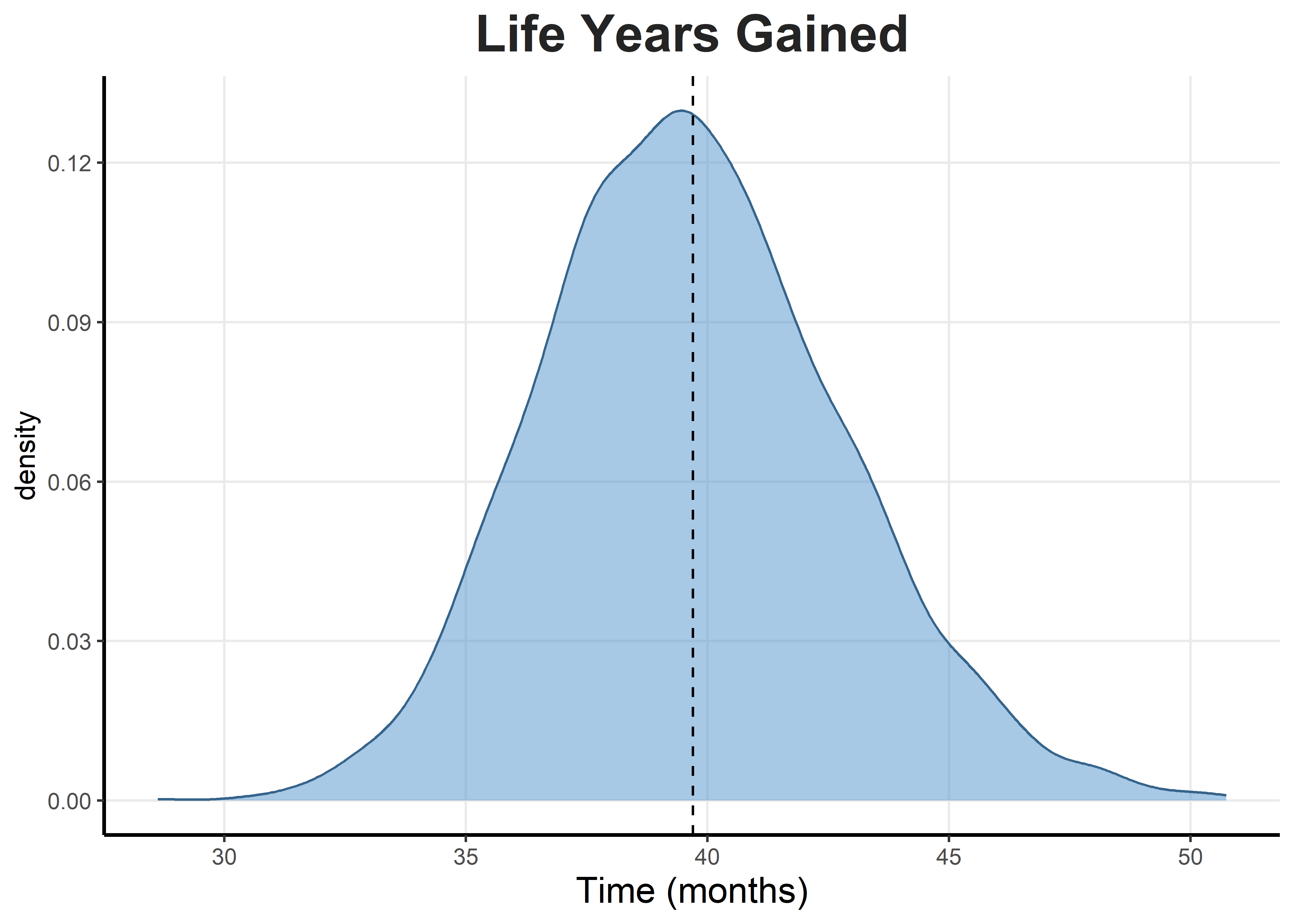}
\end{center}
\caption{LYG for the arrhythmia data. Left: LYG illustrated as the area between the two survival curves with the end time of follow-up noted by a vertical line. Right: Density of the LYG with its mean indicated by a vertical line.}
\label{fig:ca_lyg}
\end{figure}

\section{Discussion}
\label{sec:discuss}

Fitting independent poly-hazard models to the external population and the disease groups can give seemingly good fit during the follow-up period at the cost of unstable and/or unreliable  extrapolation. This article uses methods that can avoid such erratic behaviour by transferring appropriate information from data external to the disease group. The external population is based upon mortality projections as opposed to past data and the evidence is synthesized in a Bayesian context. We found that modeling on the hazard scale can identify incoherent fit more efficiently compared to the survival and cumulative hazard scale. We employ an extension of the standard poly-hazard models introducing change-points, offering added flexibility.

A review of tests for crossing survival curves may be found at \citet{li2015statistical}. Our work is concerned with a more involved issue, not simply testing for equality, but fully modeling the survival curves and the calculation of their difference, as necessitated in health economic evaluations. To do so, we transfer long term evidence to the current context thus facilitating relatively stable extrapolation. Alternatives to poly-hazard models include splines (\citealt{jackson2023survextrap}) and fractional polynomials (\citealt{royston2002flexible}). Both approaches offer nearly arbitrary flexibility at the cost of typically limited interpretability of their parameters. Perhaps more importantly, poly-hazard models offer a natural balance between efficient use of the data and undesirable carry-over effects.

This work was motivated by 3 use cases that include a crossing survival problem, an emerging cancer therapeutic and a competing risks example. The unified approach presented here naturally caters for survival extrapolation in these cases. We estimate that women with breast cancer lose on average 10.17 years compared to the general population and we also estimate the additional deficit of the triple negative sub-population. The mRNA-4157 (V940) therapeutic in combination with pembrolizumab is currently being assessed for treatment of melanoma and by incorporating the early trial evidence within our framework we estimate that an average of 3.64 extra years of life may be gained compared to pembrolizumab alone. The cardiac arrhythmia estimates mortality reduction when switching from AAD to ICD treatment by retaining a common hazard on death from other causes. 

The results are assessed under five extrapolation methods. The baseline method performs well in all our examples by appropriately modelling the bond between the disease and external populations. The constant difference and ratio methods build upon the findings of the follow-up period and extrapolate the last observed relationship between the two populations into the future. We further amend these two methods by applying them to the components of interest alone and obtain the cause-specific analogues. The proposed methodology does require individual-level data. This problem can be resolved by digitizing published Kaplan-Meier curves as illustrated in the melanoma and cardiac arrhythmia data. Still, some inaccuracies may result and care is required in this aspect. In addition, expert opinion may be elicited as done by \citet{benaglia2015survival} for the age-specific proportions of arrhythmia.

Survival analysis is also used in non-medical application areas of statistics, including industrial, actuarial and reliability functions where the idea of transferring information from an external source can be fruitful if extrapolation is required. Further development of the piecewise poly-hazard models may be considered by training them based upon alternative combinations of the disease and external groups. The predictive ability of the models developed in this article has been informally assessed, since in reality it is dictated by the performance on future/unseen data. One alternative is to characterize the model's properties using an appropriate loss function on the combined clinical and registry data and this represents the subject of current work.

\section*{Acknowledgements}

The authors are grateful to Dr. Pippa Corrie for comments on the melanoma example.

\bigskip
\begin{center}
{\large\bf SUPPLEMENTARY MATERIAL}
\end{center}

\begin{description}

\item[A. The details of KM digitization] The procedure of digitizing a curve from a given plot is described.

\item[B. Some comments on typical extrapolations] Comments regarding the relation between extrapolation methods are provided.

\item[C. Mortality projections] Figures regarding the mortality analysis used.

\item[D. Precision breast cancer results] Results on the extrapolation of breast cancer data related with the 3N and n3N groups.

\item[E. Breast cancer results] Results on the application of breast cancer.

\item[F. Advanced melanoma results] Results on the application of advanced melanoma.

\item[G. Cardiac arrhythmia results] Results on the application of cardiac arrhythmia.

\end{description}

\bibliographystyle{Chicago}

\bibliography{bibtex}
\end{document}